\documentclass[prl,twocolumn,twoside]{revtex4}
\usepackage{graphicx}
\usepackage{amsmath}
\begin{document}
\title{The Gravity Tunnel in a Non-Uniform Earth}
\author{Alexander R. Klotz\footnote{klotza@physics.mcgill.ca}}
\affiliation{Department of Physics, McGill University}
\begin{abstract} How long does it take to fall down a tunnel through the center of the Earth to the other side? Assuming a uniformly dense Earth, it would take 42 minutes, but this assumption has not been validated. This paper examines the gravity tunnel without this restriction, using the internal structure of the Earth as ascertained by seismic data, and the dynamics are solved numerically. The time taken to fall along the diameter is found to be 38 rather than 42 minutes. The time taken to fall along a straight line between any two points is no longer independent of distance, but interpolates between 42 minutes for short trips and 38 minutes for long trips. The brachistochrone path (minimizing the fall time between any two points) is similar to the uniform density solution, but tends to reach a greater maximum depth and takes less time to traverse. Although the assumption of uniform density works well in many cases, the simpler assumption of a constant gravitational field serves as a better approximation to the true results.\end{abstract}
\maketitle
\section{Introduction}

The idea of the gravity tunnel was proposed by Cooper in 1966 in the American Journal of Physics \cite{cooper}. He showed that a tube drilled straight through the Earth along its diameter would take 42 minutes to fall through, given some assumptions and ignoring various engineering considerations. In addition, he showed that a straight tube connecting any two points could be traversed in the same amount of time, independent of distance, and the time could be made shorter with a more efficient path. In a subsequent issue, five technical comments on Cooper's original paper appeared. Kirmser \cite{kirmser} lamented the insufficient literature review and pointed out that the idea can be found in an 1898 engineering textbook, although the concept appears in an 1883 French magazine \cite{redier}. Venezian \cite{venezian}, Mallett \cite{mallett}, and Laslett \cite{laslett} each derived an expression for the brachistochrone through the Earth, the path over which the total transit time is minimized. Cooper himself wrote a comment addressing these papers \cite{cooper2}, and suggested that the length-independence of cord fall times was a coincidence based on the assumption of uniform density. Some applications of the idea were discussed in subsequent issues \cite{lee} \cite{prussing}, but to this author's knowledge there has been no relaxation of the uniform density assumption.

Since it is unlikely that such a tunnel will ever be excavated in the near future, the concept serves largely a pedagogical role. In introductory physics, the diameter-length gravity tunnel is used to demonstrate the power of simple harmonic motion: it is much easier to derive the period of oscillations than to solve kinematical equations with a changing acceleration. In advanced mechanics, it is revisited as a problem of variational calculus: what is the path connecting two points that would take the shortest amount of time to fall through? 


The key assumption made when discussing the gravity tunnel is that the density of the Earth is uniform throughout. This allows the gravitational field to be linear with respect to radial position, dictating the falling object undergo simple harmonic motion. This paper examines the gravity tunnel without this assumption, to quantify its validity to obtain more accurate estimates by studying the system in greater detail.

The internal structure of the Earth is described by the Preliminary Earth Reference Model (PREM), based on reconstructions from seismic data \cite{prem}. The radial density profile can be used to reconstruct the radial mass and gravity profiles. A reproduction of PREM data can be seen in Figure 1. The Earth is denser towards the center (reaching 13 tonnes per cubic meter), and exhibits a sharp discontinuity in the density at the boundary of the outer core, dropping by nearly 50 percent. Because of this sharp discontinuity in the density, the gravitational field strength actually \textit{increases} below the surface, reaching a maximum of about 1.09 g, before decreasing in a roughly linear matter through the core to the center. 

Using the reconstructed gravitational field strength inside the Earth, three versions of the gravity tunnel are analyzed: falling through the center of the Earth to the other side, falling along a straight line dug between two points, and the brachistochrone path that minimizes travel time. Numerical integration is used as a method to generate solutions to these problems.

Although the typical assumption is that the density of the Earth is uniform, an even simpler approximating assumption can be made: that the gravitational field is constant in magnitude throughout the interior of the planet, always pointing towards the center at 9.8 N/kg. In addition to disagreeing with PREM data there are two other reasons it is unphysical: it implies a singular density at the origin, and it leads to a sharp discontinuity in acceleration as the falling object passes the origin. Nevertheless, it will become clear that this assumption works quite well at matching the realistic case.


\begin{figure}
	\centering
		\includegraphics[width=0.50\textwidth]{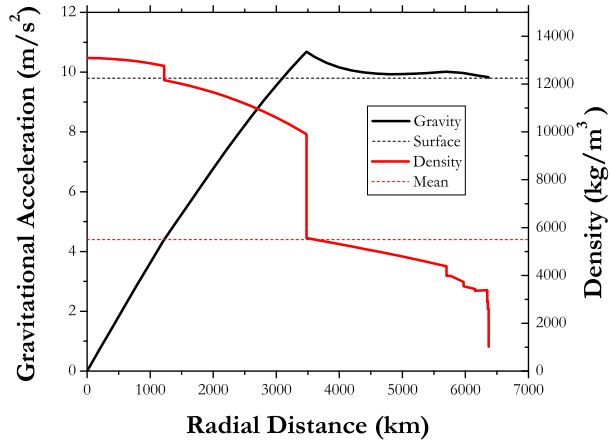}
	\label{fig:prem}
	\caption{The gravitational field strength (black) and density (red) as a function of radius inside the Earth according to the PREM \cite{prem}.}
\end{figure}

\section{Derivations and Calculations}

\subsection{A. Falling Through the Center of the Earth}

The time taken to fall through the Earth along its diameter is usually calculated under the assumption that the Earth is of uniform density $\rho$. Under this approximation, the force of gravity $F_{G}$ acting on a test mass $m$ at radial position $r$ comes from the mass of the sphere below the object, due to the shell theorem:
\begin{equation}
F_{G}(r)=-\frac{Gm\left(\frac{4}{3}\pi\rho r^{3}\right)}{r^2}=m\frac{d^{2}r}{dt^{2}}
\label{eq:force}
\end{equation}

$G$ is Newton's constant. Because the gravitational force is linear with respect to radial position, the dynamics of the falling object can be described by simple harmonic motion with an angular frequency:

\begin{equation}
\omega=\sqrt{\frac{4\pi}{3}G\rho}=\sqrt{\frac{g}{R}}\rightarrow T_{\rho}=\frac{\pi}{\omega}\approx 42 \,m
\label{eq:omega}
\end{equation}

Where $g$ is the gravitational acceleration and $R$ is the radius of the Earth. The period of these oscillations, given the average density of the Earth (5500 $kg/m^3$), is 84 minutes, meaning it would take 42 minutes to fall through a uniform Earth. The peak velocity at the center of the Earth is near 8 km/s, over thirty times the speed of a typical transatlantic aircraft. The period of oscillation and the peak velocity are the same for a circular orbit at Earth's surface.

If the gravitational field is constant inside the Earth, the time taken to fall through the Earth can be found by simple kinematics:

\begin{equation}
T_{g}=2\sqrt{\frac{2R}{g}}\approx 38\,m
\label{eq:gfall}
\end{equation}

This is shorter than the uniform density time by a factor of $\pi/\sqrt{8}$ or about 11 percent. To calculate the time taken to fall to the center of the Earth given the PREM radial gravity profile (Figure 1), the kinematical equation is integrated numerically, where the gravitational acceleration at any radial position is calculated by linear interpolation between the two closest PREM reference points. The time taken to fall through a tunnel through the center of the real Earth is almost exactly what it would be if the gravitational field were uniform throughout the Earth (Figure 2), about 38 minutes. At any given time during the fall, the position of an object falling through the Earth would be very close to its position as predicted by $R-1/2gt^2$, with deviations of up to 50 meters. To the nearest second, the fall times are: PREM: 38 minutes 11 seconds; constant gravity: 38 minutes 0 seconds; uniform density: 42 minutes 12 seconds.

\begin{figure}
	\centering
		\includegraphics[width=0.50\textwidth]{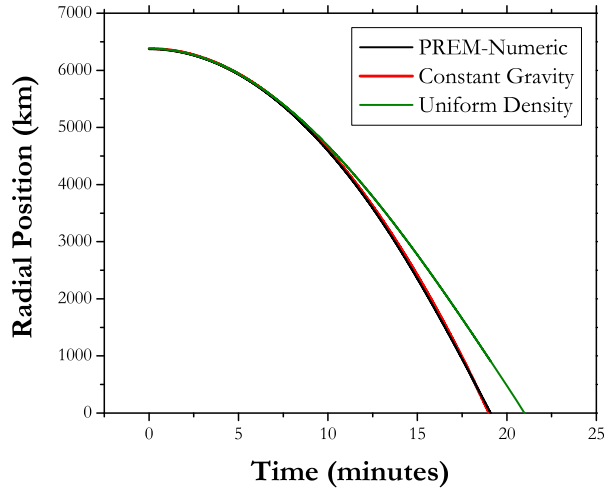}
	\label{fig:fall}
	\caption{Radial height versus time when falling to the center of the Earth according to the PREM and according to the uniform density and constant gravity approximations. The PREM curve is not identical to the constant gravity curve, but the differences are difficult to distinguish by eye.}
\end{figure}

\subsection{B. The Cord Path}

The uniform-density gravity tunnel has an interesting property that any cord path, a tunnel along a straight line between any two points on the surface, can be traversed in the same amount of time. Does the non-uniform density of the planet still preserve this feature? To answer this question we consider the kinematics of an object falling on a cord path under a central gravitational pull, according to the coordinate system in Figure 3.

\begin{figure}
	\centering
		\includegraphics[width=0.3\textwidth]{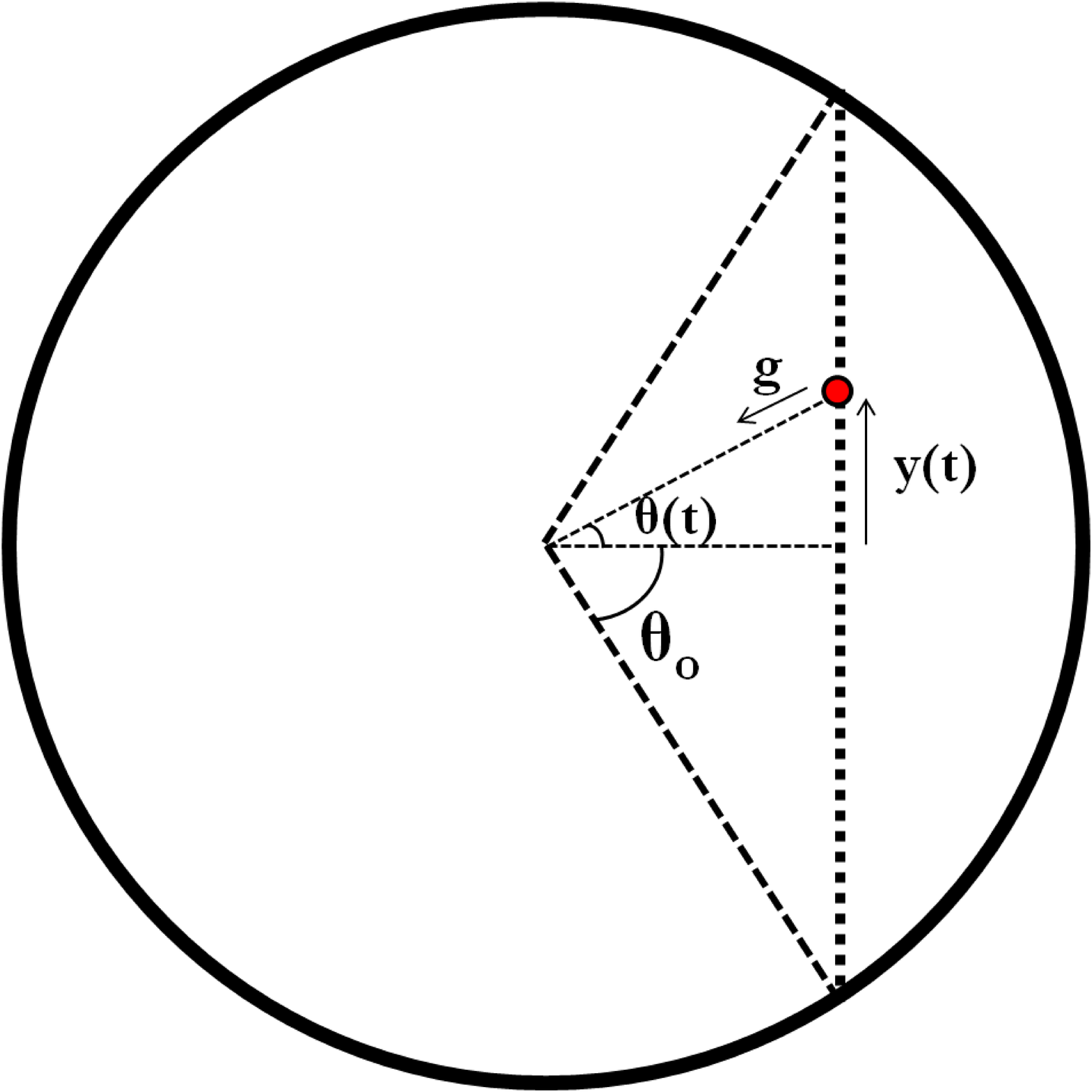}
	\label{fig:linediag}
	\caption{Diagram of the coordinate system used to find the time taken to fall through a cord path.}
\end{figure}

An object falling along a non-central linear path under the influence of central gravity experiences an acceleration:

\begin{equation}
a=\frac{d^{2}y}{dt^{2}}=-g\,sin(\theta)\hat{y}
\label{eq:aline}
\end{equation}

The position at any time can be written as:

\begin{equation}
y=R\cos(\theta_{o})\tan(\theta)
\label{eq:position}
\end{equation}

Taking the second derivative of (\ref{eq:position}) and equating it to (\ref{eq:aline}) yields a second order differential equation for the motion of an object falling along a cord path:

\begin{equation}
R\cos(\theta_{o})\left(1+\tan^{2}(\theta)\right)\left(2\tan(\theta)\dot{\theta}^{2}+\ddot{\theta}\right)=-g\sin(\theta)
\label{eq:line}
\end{equation}

The gravitational acceleration in (\ref{eq:line}) can either be the constant surface gravity, the realistic internal gravity as obtained from the PREM data, or the radially-linear gravity $g\frac{r}{R}$ from the uniform density assumption. The time taken to fall to middle of a linear path can be found numerically by integrating (\ref{eq:line}) from $\theta_{o}$ to zero using the Runge-Kutta method and recording the number of time steps in the integration.

\begin{figure}
	\centering
		\includegraphics[width=0.50\textwidth]{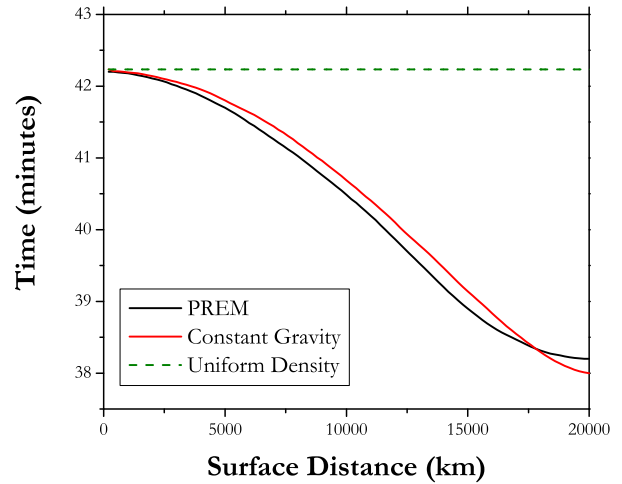}
	\label{fig:line}
	\caption{Time taken to fall through a cord path between two points at a given distance along the surface.}
\end{figure}

The cord path fall times as a function of surface distance can be seen in Figure 4. The numerical scheme replicates the distance-independence for the uniform density case, but the PREM time decreases with increasing path length, from near 42 minutes for short paths where the gravity does not deviate much from its surface value, to 38 minutes as it approaches the diameter-length fall. The constant gravity solution is again similar to the PREM solution, following the same trend with respect to distance. An exact solution exists for the constant gravity cord fall time that can be expressed in terms of elliptic integrals (see appendix), or as a Taylor series about $T_{\rho}$. The series is similar to the correction to the simple pendulum period for the initial angle dependence, the main difference being that the leading-order correction is negative. The time can be approximated well at next-to-leading order:

\begin{equation}
T_{g}(\theta_{o})=\pi\sqrt{\frac{R}{g}}\left(1-\frac{1}{16}\theta_{o}^{2}+\frac{19}{3072}\theta_{o}^{4}+O(\theta_{o}^{6})\right)
\label{eq:cordtaylor}
\end{equation}

\subsection{C. The Brachistochrone}

The brachistochrone (from the Greek for ``shortest time'') is the path that takes the least amount of time to fall between two points. The simple brachistochrone in a uniform vertical gravitational field was issued as a challenge by Johann Bernoulli in 1696, leading to the development of variational calculus. The brachistochrone path for a gravity tunnel inside a uniform Earth was considered numerically by Cooper \cite{cooper} and solved analytically by Venezian and others \cite{venezian}. Here, the brachistochrone path for an arbitrary spherical mass distribution will be derived, in order to numerically find the path through the Earth according to the PREM. 

Conservation of energy ($E$) dictates that the sum of kinetic and gravitational potential energy is constant. For an arbitrary radial mass profile $M(r)$ we have:

\begin{equation}
E=\frac{1}{2}mv^{2}+\frac{GmM(r)}{r}
\label{eq:cons}
\end{equation}

Velocity $v$ and radial position $r$ are related by the fact that the velocity is zero at the surface $r=R$:

\begin{equation}
v(r)=\sqrt {2G \left( {\frac {M(R)}{R}}-{\frac {M \left( r \right) }{r}} \right) }
\label{eq:v}
\end{equation}

The brachistochrone path between points $A$ and $B$ is found by minimizing the time integral $T$ in polar coordinates:

\[T=\int_{A}^{B}\frac{ds}{v}=\int_{0}^{\theta_{AB}}\frac{\sqrt{dr^{2}+r^{2}d\theta^{2}}}{v}\]

\begin{equation}
=\int_{0}^{\theta_{AB}}\sqrt{ \frac{r'^{2}+r^{2}}{2G \left( {\frac {M(R)}{R}}-{\frac {M \left( r \right) }{r}} \right) }}d\theta
\label{eq:time}
\end{equation}

The integrand $f$ does not explicitly depend on the angle $\theta$, so the integral can be minimized using Beltrami identity:

\begin{equation}
f-r'\frac{df}{dr'}=C
\label{eq:beltrami}
\end{equation}

Where $C$ is a constant. Evaluating, we have:

\begin{equation}
\frac{dr}{d\theta}=\sqrt{{\frac { \left( {r}^{3}R-2\,{C}^{2}GM(R)r+2\,{C}^{2}GM \left( r
 \right) R \right) {r}^{2}}{2G \left( M(R)r-M \left( r \right) R \right) {
C}^{2}}}}
\label{eq:slope}
\end{equation}

Because the slope of the path is flat when at its maximum depth, a relationship between $C$ and $dr/d\theta$ can be found when $r=R_d$:

\begin{equation}
C=\sqrt {\frac {R{R_{{d}}}^{3}}{2G \left( M(R)R_{{d}}-M \left( R_{{d}} \right) R \right) }}
\label{eq:const}
\end{equation}

Without knowing the radial mass profile, equations (\ref{eq:slope}) and (\ref{eq:const}) define a general brachistochrone. A simplified mass profile simplifies the expressions. For the case of uniform density, we have:

\begin{equation}
\frac{dr}{d\theta}=\frac{rR}{R_{d}}\sqrt {{\frac {{r}^{2}-{R_{d}}^{2}}{{R}^{2}-{r}^{2}}}}
\label{eq:rhobrac}
\end{equation}
For the case of constant gravity, we have:
\begin{equation}
\frac{dr}{d\theta}=\frac{r}{R_{d}}\sqrt {{\frac { \left( r-R_{d}\right)  \left( r \left( R-R_{{d}}
 \right) +RR_{{d}} \right) }{R-r}}}
\label{eq:gbrac}
\end{equation}

The known analytic solution to Equation \ref{eq:rhobrac} is that of a hypocycloid curve: the shape traced by a circle of diameter $(R-R_{d})$ rolling inside a circle of radius $R$. Equation \ref{eq:gbrac} is similar to the classic brachistochrone proposed by the Bernoullis, but the polar geometry makes it more difficult to solve. The solutions to these models are discussed in the appendix.

Equation \ref{eq:slope} was solved numerically with the following scheme: starting with $r=R_{d}+0.01$ (to allow a nonzero derivative) and $\theta=0$ the path was calculated by Euler integration using the known derivative. The mass at a given radius was based on a linear interpolation between the two closest PREM points. The radius was increased until it reached or exceeded the total radius of the Earth. The time taken to fall through this path was calculated by solving the time integral, again with Euler integration, using the calculated values for $r$ as a function of $\theta$. The same procedure was repeated for the uniform density case (\ref{eq:rhobrac}) and the constant gravity case (\ref{eq:gbrac}). As a validation of this scheme, the times calculated for the uniform density case as a function of distance can be compared to the known analytical solution (see appendix).

\begin{figure}
	\centering
		\includegraphics[width=0.50\textwidth]{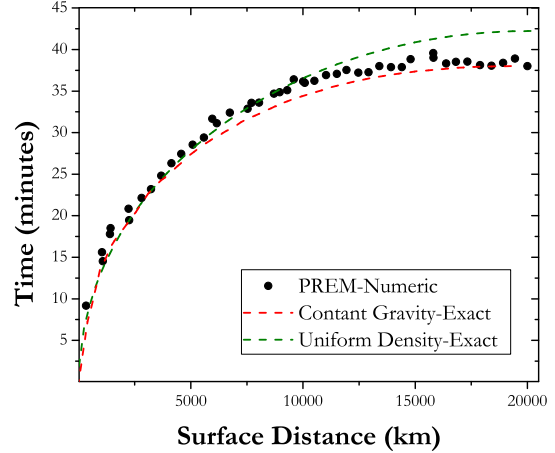}
	\label{fig:brac}
	\caption{Minimum time taken to fall through a tunnel connecting two points at distance along the surface. The numerical points have been smoothed with a rolling average of five points.}
\end{figure}

The time taken to traverse brachistochrone paths as a function of surface distance can be seen in Figure 5. In all cases, the time taken to traverse the brachistochrone path increase from zero to the diameter-length distance, be it 38 or 42 minutes, and the relationship between surface distance and time was similar. The paths connecting two points separated by about 13,000 km (approximately the distance between New York and Hong Kong) can be seen in Figure 6. The PREM path runs deeper than the uniform density path, while the constant gravity path is again similar to the PREM path. Interestingly, in some cases of the PREM model there exists two classes of solutions to the brachistochrone problem: one that follows a direct curve similar to the hypocycloid, and one that skirts the core, staying at near-constant radius at maximum depth. The core-skirting path represents a local minimum in fall time, the deeper path is faster.

\begin{figure}
	\centering
		\includegraphics[width=0.4\textwidth]{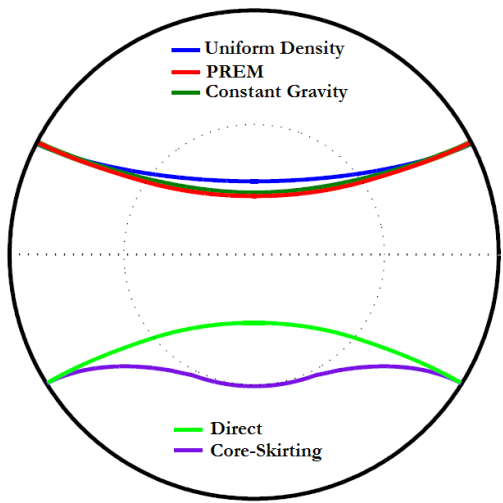}
	\label{fig:bracpath}
	\caption{Diagrams of the brachistochrone paths through the Earth. Northern hemisphere: the uniform density path does not go as deep as the PREM path, while the constant gravity path is similar. Southern hemisphere: for certain distances, two classes of solution exist for the brachistochrone problem. The direct path is faster than the one that skirts the core}
\end{figure}

\section{Discussion}

It is clear from the results presented here that the assumption that the gravitational field is uniform inside the Earth is more appropriate than the assumption that the density inside the Earth is uniform. Why does this assumption work so well? Heuristically, the gravitational field strength inside the Earth does not deviate far from its surface value until more than half-way towards the center. By the time the falling object reaches these weaker gravitational fields, it is travelling sufficiently fast that the time spent in these regions is minimized: most of the time spent falling occurs in regions where the acceleration is close to $g$. Empirically, a constant gravitational field would require that that the mass enclosed within a given volume is quadratic with radius, in order to exactly cancel Newtonian gravitation. A power law fit to the PREM mass profile for $M(r)=M\left(\frac{r}{R}\right)^{\alpha}$ yields a scaling exponent $\alpha=1.97 \pm 0.02$, very close to 2.

There are other considerations besides the uniform density that have been ignored. Regarding the diameter fall time, assuming uniform density yields an error of about 11 percent from the PREM solution, but other factors do not make as large a difference. The asphericity of the Earth, leading to differences in the radius and gravitational acceleration with respect to latitude \cite{geodetic}, only amounts to a correction of about ten seconds. The rotation of the Earth, again imparting latitude dependence, is insignificant when the fall times are much less than a day and amount to about four seconds over a forty minute trip \cite{cooper}.

The core-skirting numerical solution to the brachistochrone problem represents the fastest path that terminates at that depth, but not the fastest path connecting two points on the surface. This is an artefact of the numerical procedure, which integrated the path from the deepest point to the surface. The existence of these paths may serve as a useful alternative given that it may be difficult to excavate a tunnel through the liquid outer core. This family of paths also raises an interesting question when introducing variational calculus: how does the path ``know'' to stay at that depth before ascending without ``knowing'' the gravitational potential above or below it, or sampling every possible path. The answer lies in the fact that the local rather than global travel time is minimized between, one step of $ds$ and the next. The existence of a global and a local minimum further highlights this. The numerical methods used to study the terrestrial brachistochrone can be applied to other celestial bodies with known internal structure, and may prove useful for the future exploration of gas giants.


In his second paper \cite{cooper2}, Cooper discusses whether it is coincidental that the all linear paths (as well as the orbit half-period) take the same time. He argues that it is coincidental based on his assumption that Earth's density is uniform. Relaxing that assumption, it is seen that this equivalence is now merely a similarity: cord fall times vary by up to 11 percent from the surface orbit time. Cooper surmised that the true fall time would be ``very much different'' given a non-uniform density, but in light of the analysis presented in this paper, the answer appears closer to his approximation than he thought.

A question central to this paper is whether the assumption of uniform density is justified. The results show that it works very well at approximating the PREM solution: deviations rarely exceed ten percent. However, if simplifying assumptions are to made, it has been shown that the assumption of constant gravity is a better one. Pedagogically, this invalidates the gravity tunnel as an introductory problem for simple harmonic motion, as the fall is described by basic kinematics. The brachistochrone curve and time under this assumption may be solvable, but it is more challenging than the uniform density assumption. These hallmark textbook problems should not necessarily change in light of these computations, although a discussion of the validity of the assumptions may be appropriate.

\section{Conclusion}

Numerical analysis has been used to study the dynamics of an object falling through a gravity tunnel without the assumption that the Earth is uniformly dense. It was found that the assumption of a constant gravitational field serves as a better approximation to the PREM result than the assumption of uniform density. Overall, the longest fall times were shorter by about 11 percent. The most significant deviation from the uniform density predictions is that the fall time for a cord path is no longer independent of surface distance. The brachistochrone curves do not deviate significantly from the analytic uniform result, but there exists a class of locally minimal paths that skirt the core. These analyses show that if an assumption is to be made, it is that of constant gravity rather than of uniform density, but the latter remains useful pedagogically.

\bibliography{earthrefs}
\newpage\clearpage\raggedbottom
\begin{widetext}

\section{Appendix: Solutions to simplified brachistochrones}

This appendix discusses closed-form solutions to the brachistochrone curves under constant gravity and uniform density. The computer algebra software package Maple 9 was used to find many of these expressions. Throughout this section, $\Phi_F$ is the incomplete elliptic integral of the first kind, $\Phi_K$ is the complete elliptic integral of the first kind, $\Phi_E$ is the incomplete elliptic integral of the second kind, $\Phi_G$ is the complete elliptic integral of the second kind, $\Phi_\Pi$ is the incomplete elliptic integral of the third kind, and $\Phi_\Psi$ is the complete elliptic integral of the third kind. 
 
The constant gravity brachistochrone is traversed in time:

\[T=\int{\frac{\sqrt{dr^{2}+r^{2}d\theta^{2}}}{\sqrt{2g(R-r)}}}\]

With the minimization condition dependent on the radius at maximum depth:

\[\frac{dr}{d\theta}=\frac{r}{R_{d}}\sqrt {{\frac { \left( r-R_{d}\right)  \left( r \left( R-R_{{d}}
 \right) +RR_{{d}} \right) }{R-r}}}\]
 
The path can be found by isolating $d\theta$ in the above expression and integrating.

\begin{align*}
\theta(r)=2\,\sqrt {-{\frac {1}{R_{{d}} \left( 2\,R-R_{{d}} \right) }}} \left( 
\left( R-R_{{d}} \right)\Pi \,  +R_{{d}}F \right) 
\end{align*}
 
Where
\begin{align*}\Pi=\Phi_{\Pi} \left( \sqrt {{\frac {rR-rR_{{d}}+Rd}{{R}^{2}}}},{
\frac {R}{d}},\sqrt {{\frac {{R}^{2}}{d \left( 2\,R-d \right) }}}
 \right)&\ \ \  F=\Phi_{F} \left( \sqrt {{\frac {rR-rR_{{d}}+RR_{{d}}}{{R}^{2}}}}
,\sqrt {{\frac {{R}^{2}}{R_{{d}} \left( 2\,R-R_{{d}} \right) }}}
 \right) 
\end{align*}

Because the paths are vertical at the surface, the derivative is undefined. The distance traversed, $S$ for a given $R_{d}$ is the limiting value of $\theta(r)$:

\begin{align*}S=\lim_{r \to R}\theta(r)=\frac{2R}{\sqrt{d(2-d)}}\left(\Phi_{\Psi} \left( {d}^{-1},\sqrt {-{\frac {1}{d \left( -2+d
 \right) }}} \right)  \left( 1-d \right) +d\Phi_{K} \left( 
\sqrt {-{\frac {1}{d \left( -2+d \right) }}} \right) \right)\end{align*}

Where $d=R_{d}/R$. Finally, the total time can be found by substituting the minimal path derivative into the time integral:

\begin{align*}
&T=2\int_{R_{d}}^{R}\frac{\sqrt{1+r^{2}\frac{d\theta}{dr}}}{\sqrt{2g(R-r)}}dr= -\left(  \left( R-R_{{d}} \right) \sqrt {-{\frac {R_{{d}}g}{2\,R-R_{{d }}}}} \left( L-K \right)+\sqrt {-R_{{d}}g \left( 2\,R-R_{{d}} \right) } \left( {\it G}-{\it E} \right) \right) \frac{\sqrt{2}}{g\sqrt{R-R_{d}}}
\end{align*}

Where:

\begin{align*}
&{\it G}=\Phi_{G} \left( \sqrt {{\frac {{R}^{2}}{R_{{d}}
 \left( 2\,R-R_{{d}} \right) }}} \right) & E=\Phi_{E} \left( \sqrt {{\frac {R_{{d}} \left( 2\,R-R_{{d}}
 \right) }{{R}^{2}}}},\sqrt {{\frac {{R}^{2}}{R_{{d}} \left( 2\,R-R_{{
d}} \right) }}} \right) \\
&{\it K}=\Phi_{K} \left( \sqrt {{\frac {{R}^{2}}{R_{{d}}
 \left( 2\,R-R_{{d}} \right) }}} \right)
&L=\Phi_{F}\left( \sqrt {{\frac {R_{{d}} \left( 2\,R-R_{{d}}
 \right) }{{R}^{2}}}},\sqrt {{\frac {{R}^{2}}{R_{{d}} \left( 2\,R-R_{{
d}} \right) }}} \right)
\end{align*}

The complex terms arising from negative roots are balanced by similar terms from the elliptic functions.

By comparison, the brachistochrone for the uniform density Earth, as derived by Venezian \cite{venezian} is simpler:

\[\theta(r)=\arctan \left( R\sqrt {{\frac {{r}^{2}-{R_{{d}}}^{2}}{{R}^{2}-{r}^{2}}
}}{R_{{d}}}^{-1} \right) -\frac{R_{{d}}}{R}\arctan \left( \sqrt {{\frac {{r}^{2}
-{R_{{d}}}^{2}}{{R}^{2}-{r}^{2}}}} \right) \]
Because the hypocycloid curve is defined by a small circle rolling inside a larger one, the ratio between surface distance and maximum depth is simply $\pi$, and the time taken to traverse the path is:
\begin{equation*}T=\sqrt{\frac{S}{R}\frac{2\pi\,R-S}{g}}) \end{equation*}

The time taken to fall along a linear path under constant gravity can be found by the same method:

\[T_{line}=\int_{-\theta_{o}}^{\theta_{o}}\frac{R\cos\theta_{o}\tan\theta}{\sqrt{2g(R-R\frac{\cos\theta_{o}}{\cos\theta})}}d\theta=\sqrt{\frac{8(1-\cos\theta_{o})}{\sin(\theta_{o})^2}{\frac{R}{g}}}\left(\Phi_{G}\left(\sqrt{\frac{1-\cos\theta_{o}}{1+\cos\theta_{o}}}\right)(1+\cos\theta_{o})+\Phi_{K}\left(\sqrt{\frac{1-\cos\theta_{o}}{1+\cos\theta_{o}}}\right)\cos\theta_{o}\right)\]

To validate the numerical scheme used to generate the PREM brachistochrone, these analytic expressions can be compared to numerical solutions with the same potential.

\begin{figure}[h]
\includegraphics[width=0.5\textwidth]{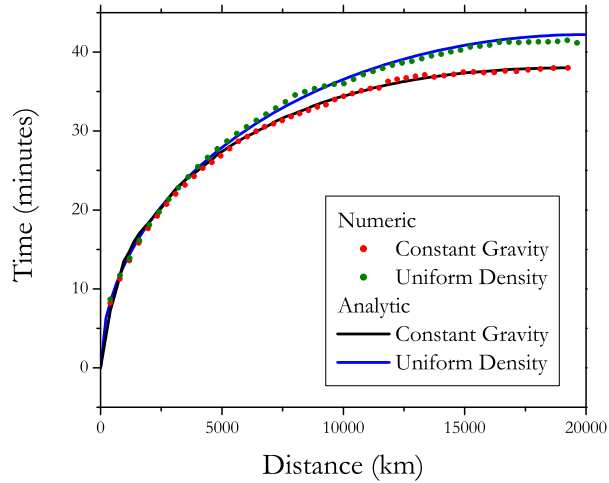}%
\label{fig:appendix}%
\caption{Comparison of analytic and numeric solutions to the brachistochrone travel time for the constant gravity and uniform density approximations.}
\end{figure}

\end{widetext}

\end{document}